# Principal component analysis model for machine-part cell formation problem in group technology

Wafik Hachicha[a], Faouzi Masmoudi[a,b] and Mohamed Haddar[a,b]

a *Unité de recherche de Mécanique, Modélisation et Production,(U2MP).*
b *Département de génie mécanique, Ecole Nationale d ingénieurs de Sfax, B.P. W, 3038 Sfax, Tunisia*

**Abstract**

In this paper, we consider the problem of forming machine cell in cellular manufacturing (CM). The major problem in the design of a CM system is to identify the part families and machine groups and consequently to form manufacturing cells. The aim of this article is to formulate a multivariate approach based on a correlation analysis for solving cell formation problem. The proposed approach is carried out in two phases. In the first phase, the correlation matrix is used as an original similarity coefficient matrix. In the second phase, Principal Component Analysis (PCA) is applied to find the eigenvalues and eigenvectors on the correlation similarity matrix. A scatter plot analysis as a cluster analysis is applied to make machine groups while maximizing correlation between elements. A numerical example for the design of cell structures is provided in order to illustrate the proposed approach. The results of a comparative study based on multiple performance criteria show that the present approach is very effective, efficient and practical

## 1. Introduction

Machine layout in a traditional production system is mainly process (functional) oriented where machines performing similar processes are grouped together. Parts requiring more than one process travel from one section of a production system to another until their operation requirements are completed. Long and uncertain throughput times are usually the major problems in such a system [1]. Group technology (GT) has been proposed as a layout approach to circumvent the above-mentioned problems.

GT is a manufacturing philosophy that has attracted a lot of attention because of its positive impacts in the batch-type production. In essence, GT tries to retain the flexibility of a job shop with the high productivity of a flow shop. GT whose basic idea is to decompose a manufacturing process into a set of subsystems for the sake of better control possesses a manufacturing philosophy that identifies and exploits the similarities of product design and manufacturing process. This characteristic of GT leads to simplified material flows, reduced material handling, reduced work-in-progress inventory, reduced throughput time, improved sequencing and scheduling on the shop floor

Cellular manufacturing (CM) is one of the applications of GT principles to manufacturing. In the design of a CM system, similar parts are grouped into families and associated machines into groups so that one or more part families can be processed within a single machine group. The process of determining part families and machine groups is referred to as the cell formation (CF) problem. CM has been recognized as one of the most recent technological innovations in job-shop or batch-type production to gain economic advantages similar to those of mass production. Many firms have recently started to adopt CM systems in order to achieve flexibility and efficiency, which are crucial for survival in today's competitive environment.

The main used techniques are classification and coding systems (such as in [2], [3] and others), machine-component group analysis, mathematical and heuristic approaches (such as in [5], [5], [6] and others), similarity coefficient based on clustering methods (such as in [7], [8], and others), graph-theoretic methods, knowledge-based and pattern recognition methods, fuzzy clustering methods, evolutionary approaches (such as in [9] and others) and neural network approaches. A number of researches have published review studies for existing CF literature (refer to [10], [11] and others). Among these techniques, those based on similarity coefficients are more basic and more flexible for dealing with the CF problem [11], [13]. Although a number of research papers have used different types of similarity and dissimilarity coefficients for identifying part families and machine cells. A similarity coefficient represents the degree of commonality between two parts or two machines. The binary data based problems consider only assignment information, that is, a part need or need not a machine to perform an operation.

The initial machine-part incidence matrix is a binary matrix whose rows are machines and columns stand for parts. Where $a_{ij} = 1$, means that machine **i** (1....m) is necessary to process part **j** (1....p) and $a_{ij} = 0$, otherwise.

Many definitions of similarity coefficient have been proposed for GT problem (such as in [4], [6], [12] and others). A clustering algorithm must transform the initial machine-part incidence matrix into the final matrix with structured form (blocks in diagonal).

We are interested in finding solutions of a CF problem, which respect the following hypothesis:
- Each machine is considered as unique: even if two machines are functionally similar, they are considered as different in the model.
- One and only one routing has to be selected for each part type.

## 2. Description of the proposed approach

This approach consists in solving machine-part grouping problem using correlation as a new definition of similarity coefficient and to use the PCA as a cluster method. These techniques allow the identification of part families and machine groups simultaneously. The proposed approach consists in two phases as mentioned in figure 1.

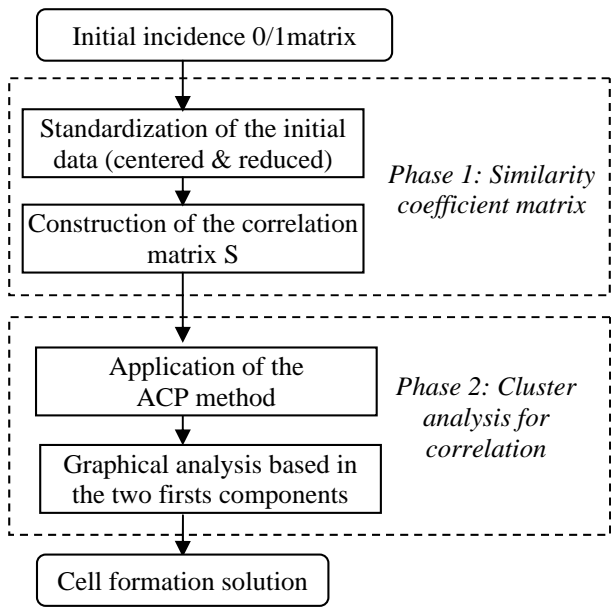

**Figure.1. Architecture of the proposed approach**

### 2.1 Phase 1: Similarity coefficient matrix

The first phase consists in building a similarity matrix. The initial machine-part incidence matrix shown in figure 2 is a binary matrix which rows are parts and columns stand for machines. Where $a_{ij} = 1$ if machine j is required to process part i and $a_{ij} = 0$ otherwise. Note that this proposed definition looks like the transpose of classical incidence matrix.

In order to explain the methodology of the proposed approach, a manufacturing system is considered with seven machines (labeled M1-M7) and eleven parts (labeled P1-P11). This example is provided by (Boctor, [6]).

Cell formation problem can be considered as a dimension reduction problem. A large number of interrelated machines are grouped into a smaller set of independent cells and a large number of interrelated parts are grouped into families. To make the initial matrix (A) more sufficiently meaningful and significant, its standardization is needed. It is expressed by [14]:

$$M_j^B = \frac{M_j^A - E_j}{\sigma_j} \quad (1)$$

where $M_j$ is a binary row vector from the matrix A:
$M_j^A [a_{1j}, a_{2j}, ....., a_{pj}]$
$E_j$ is the average of the row vector $M_j$,

$$E_j = \frac{\sum_{k=1}^{p} a_{kj}}{p} \quad (2)$$

and $\sigma_j^2 = \frac{1}{p}\sum_{k=1}^{p}(a_{kj} - E_j)^2 \quad (3)$

```
       M1  M2  M3  M4  M5  M6  M7
P1  ⎡  1   1   0   0   0   0   0 ⎤
P2  ⎢  0   1   1   0   0   0   0 ⎥
P3  ⎢  1   0   0   0   1   1   0 ⎥
P4  ⎢  0   0   0   1   0   1   0 ⎥
P5  ⎢  0   0   0   1   0   0   1 ⎥
P6  ⎢  0   1   1   0   0   0   0 ⎥
P7  ⎢  1   0   0   0   1   0   0 ⎥
P8  ⎢  0   0   0   0   0   0   1 ⎥
P9  ⎢  0   0   1   0   0   0   0 ⎥
P10 ⎢  0   0   0   1   0   0   1 ⎥
P11 ⎣  1   0   0   0   0   1   0 ⎦
```

**Figure. 2 . Initial machine-part incidence matrix**

The proposed similarity coefficient is based on the simple correlation matrix the incidence matrix. The correlation matrix S is defined as follows: $S_{ij}$ is m x m matrix which elements are given by: $S_{ii} = 1$ and

$$S_{ij} = \frac{1}{p}\sum_{k=1}^{p} b_{ik} b_{jk} \quad (4)$$

The similarity matrix S is show in figure 3. Detailed description of principal component analysis can be found in the relevant literature such as [15], [16], [17] and others.

```
       M1     M2     M3     M4     M5     M6    M7
M1  ⎡  1,00                                           ⎤
M2  ⎢ -0,04   1,00                                    ⎥
M3  ⎢ -0,46   0,54   1,00                             ⎥
M4  ⎢ -0,46  -0,38  -0,38   1,00                      ⎥
M5  ⎢  0,62  -0,29  -0,29  -0,29   1,00               ⎥
M6  ⎢  0,39  -0,38  -0,38   0,08   0,24   1,00        ⎥
M7  ⎣ -0,46  -0,38  -0,38   0,54  -0,29  -0,38  1,00  ⎦
```

**Figure 3: similarity matrix S**

### 2.2 Phase 2: Cluster analysis for correlation

In the second phase of the proposed approach, the machine groups and part families are identified using factor and graphical analysis. The objective is to find machine groups, part families and parts common machines using some classification scheme given by using Principal component analysis PCA representation of the data.

Factor analysis is a powerful multivariate analysis tool used to analyze interrelationships among a large number of variables to reduce them into a smaller set of independent variables called factors. Factor analysis was developed in 1904 by Spearman in a study of human ability using mathematical models [18]. Since then, most of the applications of factor analysis have been in the psychological field. Recently, its applications have expanded to other fields such as mineralogy, economics, agriculture and engineering. Factor analysis requires having data in form of correlations, and uses different methods for extracting a small number of factors from a sample correlation matrix. These methods include: common factor analysis, principal component analysis, image factor analysis, and canonical factor analysis. Detailed description of PCA method can be found in the relevant literature such as in [17], [18] and others.

PCA is the most widely used. It is an investigated of the data that is largely widespread among users in many areas of science and industry. It is one of the most common methods used by data analysts to provide a condensed description. PCA is a dimension reduction technique which attempts to model the total variance of the original data set, via new uncorrelated variables called principal components. PCA consists in determining a small number of principal components that recover as much variability in the data as possible. These components are linear combinations of the original variables and account for the total variance of the original data

In this application of PCA, the objective is clustering machines in group and parts in families. A binary decision is applied at each machine and part. Two principal components are enough to analyse correlation between elements (machines and parts). There should be high correlation among machines strongly associated with the same cell, and low correlation among machines that are associated with different cells.

The data can be represented by a two dimensional scatter plot (figure. 4) where each machine is represented by a line from the origin and each part is represented by a dot located at its weight in each line (machine). Four principal situations for the classification of machines can be recovered:

- Two neighbor machines which have a low angle distance measure, consequently they belong to the same cell. Examples can be illustrated in the figure 4 by ($M_4$ and $M_7$) and ($M_1$ and $M_5$).
- Two machines which the angle distance measure between them is almost 180°, this means that they are negatively correlated and mustn't be belong to the same cell.
- Two machines which the angle distance measure between them is almost 90°, this means that they independent, then they don't also belong to the same cell. Examples can be illustrated in figure 4 by ($M_1$ and $M_3$) and ($M_4$ and $M_6$).
- If no one of these three cases above is verified, the machine is affected to the more neighbor than affected other machine. In the CF literature problem, this machine is called an exceptional machine.

The same method is used for the classification of parts: when a part is close to a line (machine), it is assigned to the cell which component this machine. Example can be illustrated in the figure 4 by ($P_8$ and $M_7$). Otherwise, it is an exceptional part which can be illustrated, for example, in the figure 4 by $P_4$. In this situation, the exceptional part is affected to the more neighbor than affected other machine. The part $P_4$ was affected to machine $M_4$.

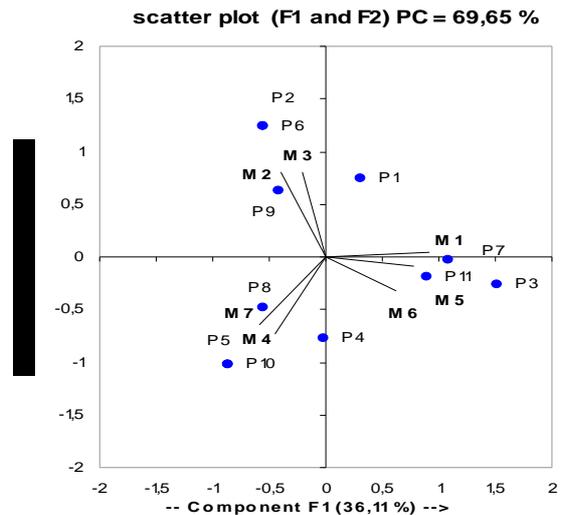

**Figure. 4. Graphical illustration of the scatter plot**

Applying the second phase of the proposed approach to these data sets yields the result shown in figure 5. We obtained the following results: The best grouping for the seven machines is to group them into tree cells: cell 1 consists of machines 2 and 3; cell 2 consists of machines 1, 5 and 6; while cell 3 consists of machines 4 and 7. In addition, cell 1 contains parts 1, 2, 6, and 9, cell 2 contains parts 3, 7, 9 and 11while cell 3 contains parts 4, 5, 8, and 10. The final solution is shown in Table 1.

**Table. 1. Final cell formation**

|     | M2 | M3 | M1 | M5 | M6 | M4 | M7 |
|-----|----|----|----|----|----|----|----|
| P1  | 1  | 0  | 1  |    |    |    |    |
| P2  | 1  | 1  |    |    |    |    |    |
| P6  | 1  | 1  |    |    |    |    |    |
| P9  | 0  | 1  |    |    |    |    |    |
| P3  |    |    | 1  | 1  | 1  |    |    |
| P7  |    |    | 1  | 1  | 0  |    |    |
| P11 |    |    | 1  | 0  | 1  |    |    |
| P4  |    |    |    |    | 1  | 1  | 0  |
| P5  |    |    |    |    |    | 1  | 1  |
| P8  |    |    |    |    |    | 0  | 1  |
| P10 |    |    |    |    |    | 1  | 1  |

## 3. Computational results

### 3.1 Performance criteria

To evaluate the performance of the proposed approach, three objective criteria widely used in the literature are selected. These criteria are the percentage of exceptional elements, machine utilization, and the grouping efficiency.

**Table 2**
**Summary of proposed approach results and the best-known results using published incidence matrices**[*]

| No. | Size | N Cell | Proposed approach results | | | Best-known results | | | Reference |
|---|---|---|---|---|---|---|---|---|---|
| | | | PE | MU | GE | PE[*] | MU[*] | GE[*] | |
| 1 | 5 x 7 | 2 | 12.50 | 82.35 | 73.68 | 12.5 | 82.35 | 73.68 | [21] King and Nakornchai (1982) |
| 2 | 15 x 10 | 3 | 00.00 | 92.00 | 45.10 | 00.00 | 92.00 | 45.10 | [3] Chan and Milner (1982) |
| 3 | 8 x 20 | 3 | 14.75 | 100.00 | 85.25 | 14.75 | 100.00 | 85.25 | [19] Chandrasekharan, and Rajagopalan (1986) |
| 4 | 14 x 24 | 4 | 3.28 | 68.60 | 67.05 | 3.28 | 68.60 | 67.05 | [2] King, (1980) |
| 5 | 24 x 40 | 7 | 40.46 | 59.09 | 42.16 | 42.14 | 53.19 | 38.07 | [19] Chandrasekharan, and Rajagopalan (1986) |

[*] The best-know results found in the literature

The first is called the Percentage of Exceptional elements (PE) and defined as the ratio of the number of elements to the number of unity elements in the incidence matrix:

$$PE = \frac{EE}{UE} \times 100 \quad (5)$$

Where UE denotes the number of unity elements in the incidence (i.e. total number of operations in the data matrix).

The second criterion is called Machine Utilization (MU) and indicated the percentage of time the machines within the clusters are used in production. MU is defined by [19] as

$$MU = \frac{UE - EE}{\sum_{k=1}^{NCell} m_k p_k} \times 100 \quad (6)$$

Where $m_k$ and $p_k$ denote, respectively, the number of machines in cell k and number of parts in family k. NCell is the number of cells.

The third criterion is called Grouping Efficacy (GE) and defined by [19] as

$$GE = \frac{UE - EE}{UE + VE} \times 100 \quad (7)$$

Where VE denotes the number of voids elements in the diagonal blocks. A void indicates that a machine assigned to a cell is not required for the processing of a part in the cell (number of 0s inside the diagonal blocks).

### 3.2 Performance measure

In order to evaluate the proposed approach and to compare its performance with other cell formation methods, five sets of data (problems) have been chosen from the literature. Table 2 summarizes the results of the comparative study and the sources of these data sets, where the performance criteria without asterisks on the left denote the results from the present approach and the performance criteria with asterisks on the right denote the best-known results in the literature

Basically, the results of the proposed approach are the same as those found in recent literature ([4], [5], [7]). These recent researches were compared with former methods like Rank Order Clustering [2], Direct Clustering Algorithm [3] and others. These recent researches demonstrated to be better in comparative studies. Therefore, it could be said that the proposed approach is valid It is more flexible and able to get correlation information between each machine and part.

### 4. Final conclusion

In this paper, a new approach is presented for part-family and machine-cell formation. The main aim of this article is to formulate a correlation analysis model to generate optimal machine cells and part families in GT problems. The correlation matrix for similarity machine and part is used as similarity coefficient matrix. The objective of PCA method is clustering machines in group and parts in families. In addition it can find the optimal number of cells.

This approach has the flexibility to allow the cell designer to either identify the required number of cells in advance, or consider it as a dependent variable. Another aspect of this research, which makes it easily portable into practice, is that it uses algorithms, which are available in many commercial software packages. For example, factor analysis can be performed on most statistical packages including SPSS, SPAD, XLSTAT, S-PLUS, and others. The proposed approach has been developed to address some deficiencies in the existing cell formation methods. It remains to be seen how this approach can be extended to address other issues highlighted in the literature.

Although the present approach focuses on the compactness of formation solution only, it can readily accommodate other manufacturing information such as production volume, sequence and alternative routings. Extending the proposed approach to this direction is our interesting research perspective.

### References

[1] Singh Nanua, Divakar Rajamani, (1996) "Cellular manufacturing systems: Design, Planning and control", First edition, A textbook, ed. CHAPMAN & HALL.
[2] King, J. R., (1980). Machine-component grouping in production flow analysis: an approach using a rank order clustering algorithm. *International Journal of Production Research*, 18(2), 213–232.
[3] Chan, H., & Milner, D., (1982). Direct clustering algorithm for group formation in cellular manufacturing. *Journal of Manufacturing Systems*, 1(1), 65–67.
[4] Albadawi, Z., Bashir, H. A., Chen, M., (2005). A mathematical approach for the formation of manufacturing cell, *Computers & Industrial Engineering*, 48, 3-21.

[5] Wang, J., (2003). Formation of machine cells and part families in cellular manufacturing systems using a linear assignment algorithm. *Automatica*, 39, 1607-1615.

[6] Boctor F. F., (1991). A linear formulation of the machine-part cell formation problem. *International Journal of Production Research*, 29, 343-356.

[7] Cheng, C. H., Goh, C. H., Lee, A., (2001). Designing group technology manufacturing systems using heuristic branching rules. *Computers & Industrial engineering*, 40, 117-131.

[8] Mukattash, A., M., Adil, M. B., Tahboub, K., K., (2002). Heuristic approaches foe part assignment in cell formation. *Computers & Industrial Engineering*, 42, 329–341.

[9] Stawowy, A., (2006) Evolutionary strategy for manufacturing cell design. *The international Journal of Management Science: Omega*, 34, 1 – 18.

[10] Joines, J. A., King, R., E., Culbreth, C., T., (1996). A comprehensive Review of Production-Oriented Manufacturing Cell Formation Techniques. *Research funded by the NCSU Furniture Manufacturing and Management Center*, North Carolina State University, DDm-92-15432.

[11] Yin, Y., Kazuhiko, K., (2006). Similarity coefficient methods applied to the cell formation problem: A taxonomy and review. *International Journal of Production Economics*. 101, 329–352.

[12] Gupta, T., Saifoddini, H., (1990). Production data based similarity coefficient for machine-component grouping decisions in the design of a cellular manufacturing system. *International Journal of Production Research*, 28, 1247-1269.

[13] Seifoddine, H., Djassemi, M,. (1995). Merits of the production volume based similarity coefficient in machine cell formation. *Journal of Manufacturing Systems*, 14, 35-44.

[14] Chaea, S. S., Wardeb, W., D, (2005). Effect of using principal coordinates and principal components on retrieval of clusters. *Computational Statistics & Data Analysis.*

[15] Ledauphin, S., Hanafi, M., Qannari, E., (2004). Simplification and signification of principal components. Chemometrics and Intelligent Laboratory Systems, 74, 277–281.

[16] Delagarge, J., (2000). « Initiation à l'analyse des données » ; 3 ième édition, Ed Dunod.

[17] Labordere, A. H., (1977). « Analyse de données : application et méthodes pratiques ». Ed Masson.

[18] Rummel, R. J. (1988). Applied factor analysis. Evanston: Northwestern University Press.

[19] Chandrasekharan, M. P., Rajagopalan, R. (1986). An ideal seed non-hierarchical clustering algorithm for cellular manufacturing. *International Journal of Production Research*, 24, 451–464.

[20] Kumar, K. R., Chandrasekharan, M. P., (1990). Grouping efficacy: a quantitative criterion for goodness of block diagonal forms of binary matrices in group technology. *International Journal of Production Research*, 28(2), 233–43.

[21] King J. R., Nakornchai V., (1982). Machine-component group formation in group technology: review and extension, *International Journal of Production Research*, 20, 117–133.